\documentclass[useAMS,usenatbib]{mn2e}

\usepackage{epsfig, amsmath}

\title[Quantitative code comparison]{A test suite for quantitative comparison of hydrodynamics codes in astrophysics}

\author[E. J. Tasker et al.]{Elizabeth J. Tasker$^{1}$\thanks{E-mail: tasker@astro.ufl.edu (EJT))}, Riccardo Brunino$^{2}$, Nigel L. Mitchell$^{3}$, Dolf Michielsen$^{2}$, 
\newauthor
Stephen Hopton$^{2}$, Frazer R. Pearce$^{2}$, Greg L. Bryan$^{4}$, Tom Theuns$^{3,5}$\\
$^{1}$Department of Astronomy, University of Florida, Gainesville, FL 32611, USA\\
$^{2}$Department of Physics and Astronomy, University of Nottingham, Nottingham, NG7 2RD, UK\\
$^{3}$Institute for Computational Cosmology, Department of Physics, University of Durham, South Road, Durham, DH1 3LE, UK\\
$^{4}$Department of Astronomy, Columbia University, New York, NY 10027, USA\\
$^{5}$Department of Physics, University of Antwerp, Groenenborgerlaan 171, B-2020 Antwerpen, Belgium}
\begin{document}

\maketitle

\begin{abstract}
We test four commonly used astrophysical simulation codes; Enzo,
Flash, Gadget and Hydra, using a suite of numerical problems with
analytic initial and final states. Situations similar to the
conditions of these tests, a Sod shock, a Sedov blast and both a
static and translating King sphere occur commonly in astrophysics,
where the accurate treatment of shocks, sound waves, supernovae
explosions and collapsed haloes is a key condition for obtaining
reliable validated simulations. We demonstrate that comparable results
can be obtained for Lagrangian and Eulerian codes by requiring that
approximately one particle exists per grid cell in the region of
interest. We conclude that adaptive Eulerian codes, with their ability to place
refinements in regions of rapidly changing density, are well suited to 
problems where physical processes are related to such
changes. Lagrangian methods, on the other hand, are well suited to
problems where large density contrasts occur and the physics is
related to the local density itself rather than the local density
gradient.
\end{abstract}


\section{Introduction}


Over the last decade, numerical simulations have developed into a
cornerstone of astrophysical research. From interactions of vast
clusters of galaxies to the formation of proto-planetary discs,
simulations allow us to evolve systems through time and view them at
every angle. They are an irreplaceable test-bed of our physical
understanding of the Universe.


A number of hydrodynamics codes have been developed that are widely
used in this field and still more are being developed. Fundamentally,
they all do the same job; they solve the equations of motion to
calculate the evolution of matter through time. Whether the matter
represents a nebula for the birth of a star or a network of galaxy
clusters, the basic technique remains the same.


However, the algorithms used to solve these equations vary from code
to code and this results in differences in the resulting
data. Understanding the origin of these variations is vital to the
understanding of the results themselves; is an observed anomaly an
interesting piece of new physics or a numerical effect? As
observational data takes us deeper into the Universe, it becomes more
important to pin down the origin of these numerical artifacts. 


Additionally, it is difficult to compare results from simulations run with different codes. With observations, papers clearly state the properties of the instrument such as the diameter of the mirror and the wavelengths it is most sensitive to. While a brief description of the code is always included in theoretical papers, there exists no obvious conversion to other numerical techniques and therefore the results are more difficult for the reader to interpret. 


The problem of code comparison is not new and it is a topic that has
recently created a great deal of interest. The reason for its current
importance is a positive one; improved numerical techniques and
increased computer power have resulted in simulations reaching greater
resolutions than could have been imagined even a few years
ago. However, this high refinement comes at a price; as we start to
pick out the detail of these complex fluid flows, the physics we need
to consider gets dramatically more complicated.  This brings us to the
main question code comparison projects are trying to answer; can we
use the same tools for this new regime of problems?


A number of papers have come out that tackle this. One of the most
famous is the Santa Barbara Comparison Project \citep{Frenk1999} which
compared many of the progenitors of today's codes by running a model
of a galaxy cluster forming. Taking a different tack \citet{Thacker}
compared the performance of a dozen different implementations of a
single approach (in this case smoothed particle hydrodynamics, SPH) on standard astrophysical problems
that included the Sod shock, examining the range of outcomes that were
available from a single technique. They concluded that one of the
weaknesses of SPH was the weak theoretical grounding which allows
several equally viable formulations to be derived. 
More recent work includes
\citet{Agertz2007}, who focus specifically on the formation of fluid
instabilities, comparing the formation of Kelvin-Helmholtz and
Rayleigh-Taylor instabilities in six of the most utilised
codes. Additionally, \citet{OShea2005} has completed a direct
comparison between two particular codes ({\it Enzo} and {\it Gadget2},
see below) looking at the formation of galaxies in a cosmological
context.


All these projects give detailed insights into the differences between
the codes, but are unable to provide a quantitative measure of how
well a code performs in a particular aspect. This is especially true
of the cosmological-based tests of \citet{Frenk1999, OShea2005} where
the problem is not sufficiently well-posed for convergence onto a
single answer. \citet{Agertz2007} sets out a simpler problem and
compares it to analytical predictions, but the system is still
sufficiently complex not to have an exact solution. Additionally, no
previous comparison has attempted to quantitatively compare different
codes to one another; asking whether it is possible to obtain
identical results and with what conditions. Without this crucial piece
of information, it is impossible to fully assess a piece of work
performed by an unfamiliar code or to judge which code might be the most suited to a given problem type. This has resulted in somewhat general
comments being made about the differences between numerical techniques
which has led to many myths about a code's ability becoming accepted
dogma.


The set of tests we present in this paper are designed to tackle these
difficulties with the intention that they might become part of an
established test programme all hydrodynamical codes should attempt. We
present four problems that specifically address different aspects of
the numerical code all of which have expected `correct' answers to
compare to. The first two tests, the Sod shock tube and Sedov blast,
are both strong shock tests with analytical solutions. The third
and forth tests concern the stability of a galaxy cluster and are
primarily tests of the code's gravitational solver. For all four tests we
directly compare the codes against the analytic solution and present
an estimate of the main sources of any systematic error. 


The remainder of this paper is organised as follows: in section 2 we
give a short summary of the main features of each of the four codes we
have employed. In section 3 we deal with the Sod shock and Sedov blast
tests, setting out the initial and final states and comparing each
code against them. We repeat this exercise for a static and
translating King sphere in section 4. Finally we discuss our results
and summarise our conclusions in section 5.

\section{Description of the codes}

The two major techniques for modelling gases in astrophysics are
smoothed particle hydrodynamics (SPH) and adaptive mesh refinement
(AMR). In the first of these, the gas is treated as a series of
particles whose motion is dictated by Lagrangian dynamics. In AMR, the
gas is modelled by a series of hierarchical meshes and the flow of
material between cells is calculated to determine its evolution. There
are a variety of codes which utilise both techniques and four of the
major ones will be used to run the tests presented in the this paper,
two of which use SPH ({\it Hydra} and {\it Gadget2}) and two of which use
AMR ({\it Enzo} and {\it Flash}). 

\subsection{\it Enzo}

{\it Enzo} is a massively parallel, Eulerian adaptive mesh refinement code, capable of
both hydro and N-body calculations \citep{OShea2004, Bryan1997}. It
has two hydro-algorithms which can be selected by the user; the
piecewise parabolic method (PPM) and the Zeus astrophysical code. The
PPM solver \citep{Colella1984} uses Godunov's method but with a
higher-order spatial interpolation, making it third-order accurate. It
is particularly good at shock capturing and outflows. The Zeus method
in {\it Enzo} is a three-dimensional implementation of the Zeus astrophysical code developed by \citet{Stone1992}\footnote[1]{Note that the hydrodynamics code {\it ZEUS-3D} has been developed independently of {\it Enzo (Zeus)} and its performace is not equivalent.}. It is a simple, fast algorithm that allows large problems to be run at high resolution. Rather than Godunov's method, Zeus uses an artificial viscosity term to model shocks, a technique which inevitably causes some dissipation of the shock front. We compare both these hydro-schemes in these tests.

\subsection{\it Gadget2}

{\it Gadget2} is a massively parallel, Lagrangian, cosmological code that is publicly
available from the author's website \citep{Springel2005}. It is an N-body/SPH code that
calculates gravitational forces by means of the Tree method
\citep{Barnes1986} and is also able to optionally employ a Tree-PM
scheme to calculate the long range component of the gravitational
interactions.  In order to follow the hydrodynamic behaviour of a
collisional medium, the code uses the entropy-conserving formulation
of SPH described in \citet{Springel2003}: the main difference of this
approach with respect to the standard \citet{Monaghan1992}
formulation of SPH resides in the choice of describing the
thermodynamic state of a fluid element in terms of its specific
entropy rather than its specific thermal energy. This 
leads to a tight conservation of both energy and entropy in simulating
dissipation-free systems. 
Additionally, {\it Gadget2} employs a 
slightly modified parametrisation of the artificial viscosity
(by introducing the so called "signal velocity" as in
\citet{Monaghan01}). The user is allowed to set the strength of this
artificial viscosity for the specific problem being considered via an
input parameter, ${\rm ArtBulkVisc}$. The time stepping scheme adopted by the
code is a leap-frog integrator guaranteed to be symplectic if a
constant timestep for all particles is employed. In this work we have
exploited the possibility of using fully adaptive individual timesteps
for all the particles in the simulation, this being a standard
practice.

\subsection{\it Hydra}

{\it Hydra} is an adaptive particle-particle, particle-mesh code
combined with smoothed particle hydrodynamics \citep{Couchman95}. It has the significant
disadvantage that even though massively parallel versions exist
\citep{PC97}, the publically available version is not a parallel
implementation and so this code cannot, as released, be used for very
large simulations. Akin to {\it Gadget2}, {\it Hydra} uses an entropy conserving
implementation of SPH, but unlike {\it Gadget2}, {\it Hydra} does not have fully
adaptive individual timesteps. Although the timestep adapts
automatically from one step to the next all the particles move in lockstep.

\subsection{\it Flash}
\label{sec:flash}
{\it Flash} is a publicly available massively parallel Eulerian AMR code
developed by the Alliances Center for Astrophysical Thermonuclear
Flashes \citep{Fryxell2000}. Originally intended for the study of
X-ray bursts and supernovae, it has since been adapted for many
astrophysical conditions and now includes modules for relativistic
hydrodynamics, thermal conduction, radiative cooling,
magnetohydrodynamics, thermonuclear burning, self-gravity and particle
dynamics via a particle-mesh approach.  {\it Flash} uses the oct-tree
refinement scheme of the PARAMESH package, with each mesh block
containing the same number of internal zones. Neighbouring blocks may
only differ by one level of refinement with each level of refinement
changing the resolution by a factor of two.  The hydrodynamics are
based on the PROMETHEUS code \citep{Fryxell1989}. The input states for
the Riemann solver are obtained using a directionally split PPM solver
\citep{Colella1984} and a variable time step leapfrog integrator with
second order Strang time splitting is adopted \citep{Strang}.  This
work uses a modified hybrid FFTW based multigrid solver to solve
Poisson's equation and determine the gravitational potential at each
timestep. This results in a vast reduction in time spent calculating
the self-gravity of the simulation relative to a conventional
multigrid solver.  {\it Flash's} refinement and de-refinement criteria can
incorporate the adapted \citet{Lohner} error estimator. This
calculates the modified second derivative of the desired variable,
normalised by the average of its gradient over one cell.

\section{Shocking Tests}
\label{sec:shock}

One of the greatest differences between simulations performed now
versus those undertaken five years ago is the increasing importance of
modelling strong shocks accurately. While it has long been known that
the Universe is a violent place, with events such as supernovae,
galaxy mergers and AGN generating blasts which rip through the
intergalactic medium, simulations did not have the resolution to see
such phenomena in detail, so these sharp discontinuities were largely
ignored. Now, as we struggle to understand the effects of feedback in
galaxy formation, multiphase media are essential physics
\citep{Tasker2007, Tasker2006, Wada2007, Robertson2007}. In order to
attack such problems codes must be able to capture shocks with some
proficiency. These two problems, the Sod shock test and the Sedov
blast test, explicitly test the resolution of shock jumps and allow
comparison with exact analytical solutions.

\subsection{Riemann Shock Tube Problem}
\label{sec:sod}

\begin{figure}
\centering
\includegraphics[width=\columnwidth]{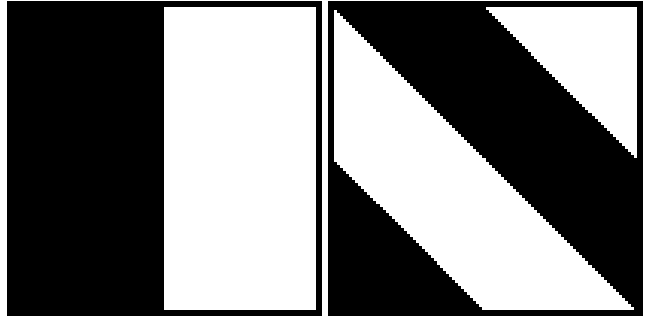}
\caption{Example projections of the initial conditions for a
  three-dimensional Sod shock test. Black and white regions represent
  fluids of different densities. The left-hand image shows the shock
  face oriented along the [1,0,0] plane, while the right-hand image
  shows it oriented along the [1,1,0] plane. In actuality our second
  test is oriented in the [1,1,1] plane i.e. oblique to all the axes.}
\label{fig:shock_setup}
\end{figure}

The shock tube problem \citep{Sod1978} has been used extensively to
test the ability of hydrodynamics codes to resolve a sharp shock
interface \citep{Feng2004, Shapiro1996, Ryu1993}. The test set-up is
simple, consisting of two fluids of different densities and pressures
separated by a membrane that is then removed. The resulting solution
has the advantage of showing all three types of fluid discontinuities;
a shock wave moving from the high density fluid to the low density
one, a rarefaction (sound) wave moving in the opposite direction and a
contact discontinuity which marks the current location of the
interface between the two fluids. For this test the initial conditions are
traditionally chosen such that the pressure does not jump across the
contact discontinuity.

We extend the traditional one-dimensional shock tube problem to
consider two three-dimensional set-ups; the first of these has the
fluid membrane at 90$^\circ$ to the x-axis of the box ([1,0,0] plane),
causing the shock to propagate parallel to this axis. In the second
test, the membrane is lined up at 45$^\circ$ to each of the $x, y$ and $z$
axes ([1,1,1] plane). This change in orientation of the shock is
designed to highlight any directional dependencies inherent in the
code, as illustrated in Figure~\ref{fig:shock_setup}. All analysis was
performed perpendicular to the original shock plane.
 
For our particular set-up we chose the initial density and pressure
jump either side of the membrane to be from $(\rho_1=4, p_1=1)$ to
$(\rho_2=1, p_2=0.1795)$ with the fluid initially at rest. The
polytropic index was $\gamma=\frac{5}{3}$. Periodic boundary
conditions were used and the results were analysed at $t=0.12$.

\subsubsection{General Results}

\begin{figure*}
\centering
\includegraphics[width=\textwidth]{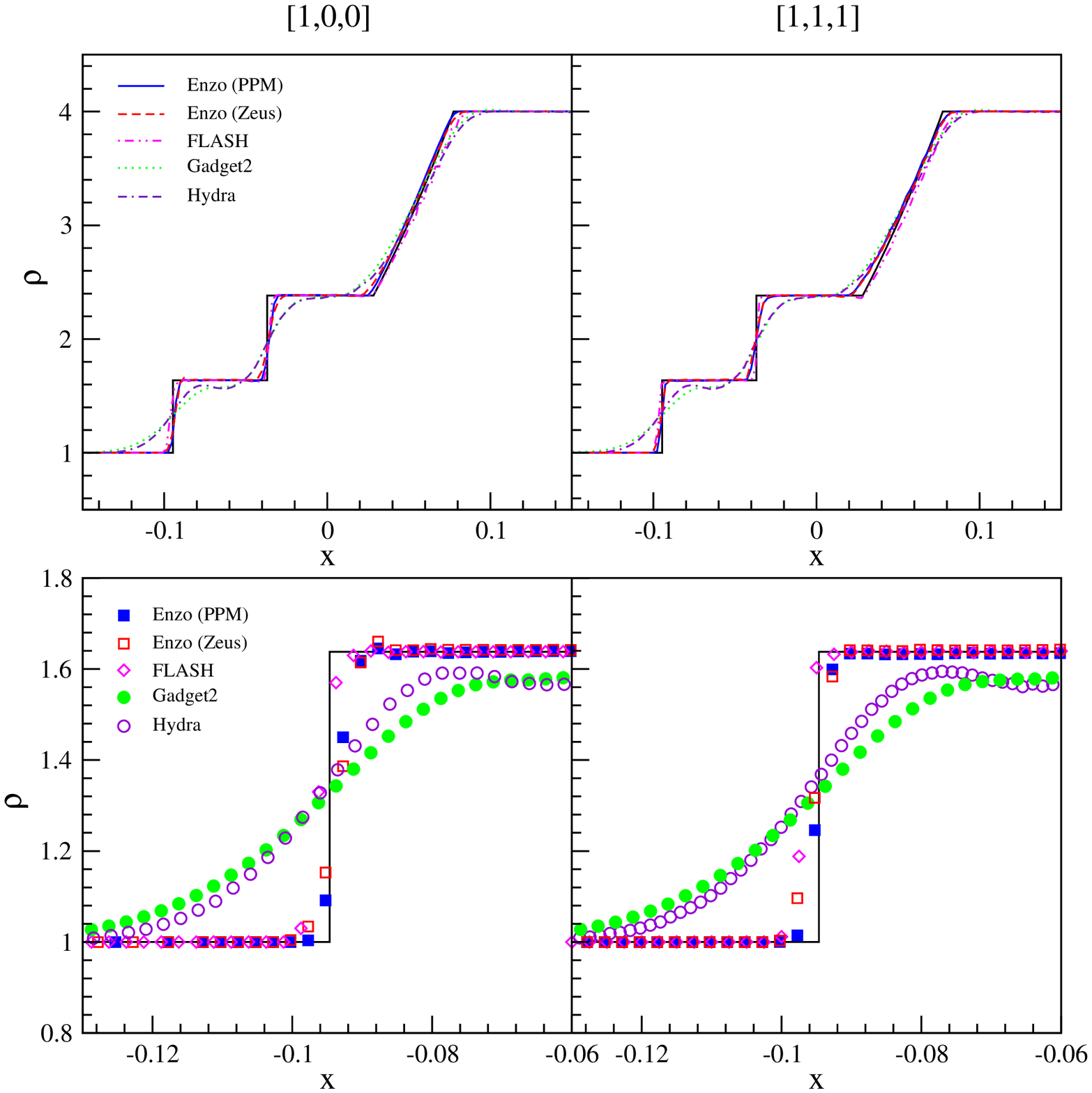}
\caption{Projected density parallel to the shock front from the
3-dimensional shock tube test. Left-hand plots show results from the 
[1,0,0] shock set up, right-hand from the [1,1,1] set-up. Both are at $t
= 0.12$. The bottom plots show a close-up of the shock front itself. }
\label{fig:sod_density}
\end{figure*}

\begin{figure*}
\centering
\includegraphics[width=11.5cm]{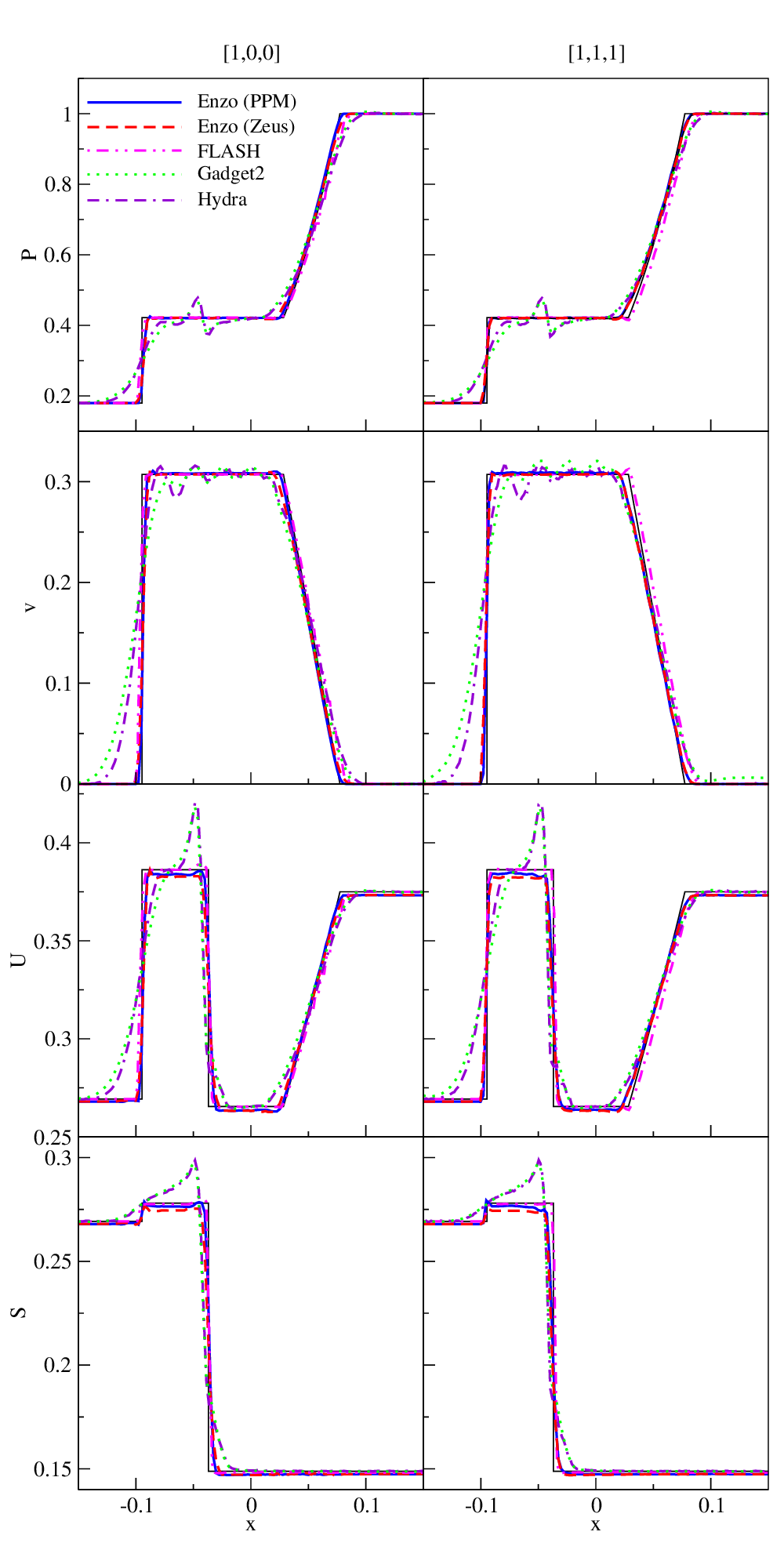}
\caption{Projected data parallel to the shock front from the
3-dimensional shock tube test. The left hand column shows the results
from the [1,0,0] shock front alignment, while the right hand column
gives those from the [1,1,1] set-up. Top-to-bottom are plotted
pressure, internal energy, velocity and entropy. As with
figure~\ref{fig:sod_density} all the panels are from $t=0.12$.}
\label{fig:sod_all}
\end{figure*}

Figures~\ref{fig:sod_density} and \ref{fig:sod_all} show the results
from all four codes running this test. In the results presented in
these figures, both {\it Enzo} and {\it Flash} used an $100^3$ initial (minimum refinement)
grid\footnote[2]{In {\it Enzo's} case, this is the `root' grid. {\it Flash} does not have a root grid, but can specify a minimum cell size that must be maintained everywhere. For ease of typing, we shall refer to the coarsest grid in both cases as the `initial' grid.} with two levels of higher refinement each of which decreased the cell
size by a factor of 2.  The smallest cell size in this case was
therefore $0.0025$ of the unit box. {\it Enzo} refined anywhere the
gradient of the derived quantities exceeded a critical value whereas
{\it Flash} placed refinements according to the \citet{Lohner} error
estimator described in section~\ref{sec:flash} which places subgrids
based on the second derivative of the derived quantities. For the SPH
codes, both {\it Hydra} and {\it Gadget2} were run with 1 million
particles formed from two glasses containing 1.6 million and 400,000
particles.  The solid black line in all cases is the analytical
solution which requires a shock at $x=-0.095$ and a contact
discontinuity in the density at $x=-0.032$ at $t=0.12$.

What is clear from Figures~\ref{fig:sod_density} and \ref{fig:sod_all}
is that all of the codes pass the zeroth level test and successfully
reproduce the shock jump conditions, although both the SPH codes
suffer from visible ringing and broadening around any
discontinuities. {\it Enzo (Zeus)} does not produce the shock jump condition as accurately as {\it Enzo (PPM)} and {\it Flash}, as seen in the plots of energy and entropy in Figure~ \ref{fig:sod_all}, where its post-shock values are around 2\% lower. In the oblique case, the quadratic viscosity term, $QV$, in {\it Enzo (Zeus)} was increased from its default value of 2.0 to 10.0. The effect of this value is discussed more fully in relation to the Sedov blast test in section~\ref{sec:code_spec}. For the planar case, $QV$ was kept at 2.0. Pleasingly none of the other codes appear to have any visible
directional dependence, performing equally well in both the grid
aligned and oblique cases we tried. All the codes could equally well
resolve the location and smooth rise of the rarefaction wave but both
the SPH codes struggle with the contact discontinuity, with a large
overshoot not seen by either of the mesh based codes. This is partly
due to the initial conditions as for the SPH codes the sudden
appearance of a density jump introduces a local source of entropy.  In
this paper we are contrasting the results from the different
approaches rather than studying the Sod shock problem itself in
detail.  As it is a standard test case higher resolution results can
be found in the individual code's method papers
\citep[e.g.][]{Fryxell2000, Thacker, Springel2005}.

Closer inspection of the data reveals differences in each code's
capacity to handle strong shocks. In the bottom row of
Figure~\ref{fig:sod_density}, we show a close-up of the density over
the shock-front. Capturing shocks accurately is an area that SPH codes
traditionally struggle with more than their Eulerian counterparts due
to their inherent nature of smoothing between particles. Indeed, we
see in this figure both {\it Gadget2} and {\it Hydra} have a smeared
out the interface compared to {\it Enzo} and {\it Flash's} steep drop in
density. Small differences between the AMR codes are also visible
here. {\it Enzo (PPM)} spreads the shock front over three cells,
whereas {\it Enzo (Zeus's)} use of the artificial viscosity term
extends this to five.  {\it Flash's} PPM scheme gives very similar
results to {\it Enzo (PPM)}, also spreading the shock front over three cells.

Figure~\ref{fig:sod_all} shows the pressure, internal energy, velocity
and computational entropy $\left(s=\frac{T}{\rho^{\gamma-1}}\right)$
over the region of interest for the planar [1,0,0] set-up (left
column) and the oblique [1,1,1] set-up (right column). 

Both {\it Hydra} and {\it Gadget2} show signs of post-shock ringing in the
velocity plot. The two lagrangian codes adopt different implementations of artificial
viscosity; with a frequently used choice of the viscosity parameter (ArtBulkVisc = 1) the ringing features in {\it Gadget2}'s profiles appears to be
more pronounced than in Hydra's (not shown in the plot). 
In order for {\it Gadget2} to get closer to Hydra's performance, 
a choice of a significantly higher viscosity parameter has been necessary
(namely ${\rm ArtBulkVisc} = 2$). 
The results produced under such a choice are shown in Figure~\ref{fig:sod_all}. The SPH codes
also exhibit a large spike in both internal energy and entropy at the
location of the contact discontinuity. This is driven by the initial
conditions, where two independent particle distributions suddenly
appear immediately adjacent to one another.

\subsubsection{Quantitative Comparison}

\begin{figure*}
\centering
\includegraphics[width=11.5cm]{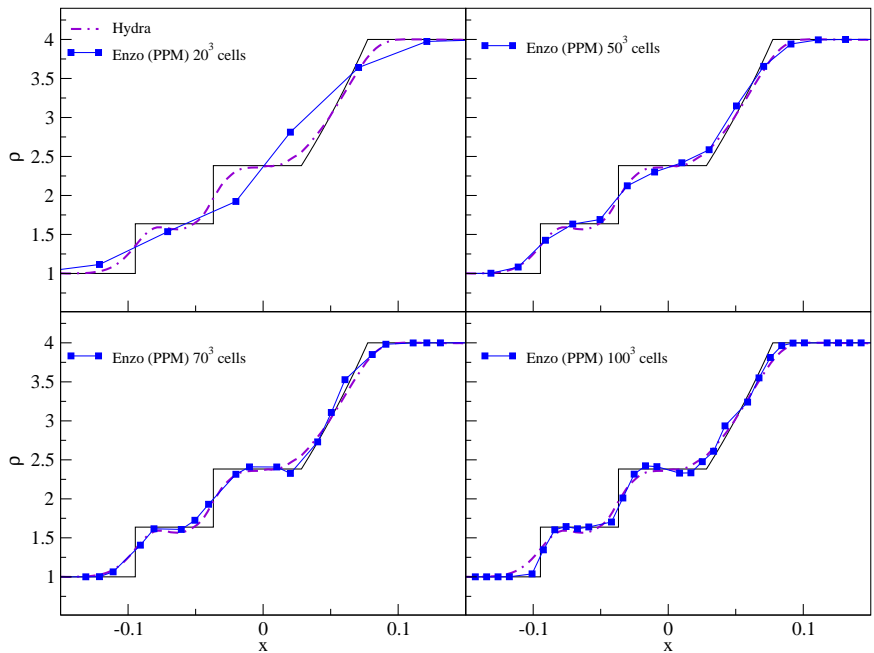}
\caption{Comparison between data from {\it Enzo (PPM)} at different
static grid sizes with {\it Hydra's} run of 1 million
particles. Results plotted show the projection of the density for the
[1,0,0] set-up.}
\label{fig:sod_comparison}
\end{figure*}

We saw in the previous section that most of the codes model shock
development and sound wave propagation with reasonable success and
that they show no directional preference to the orientation of the
shock interface. However, differences were apparent between each of
the codes, most obviously between the SPH and AMR techniques
(unsurprising, since their numerical algorithms fundamentally
differ). In this section, we quantitatively compare results from each
code and attempt to get as close a match between their results as
possible. For simplicity, we confine the AMR codes to using static
meshes for this comparison.

Figure~\ref{fig:sod_comparison} shows a graphical comparison of the
density projection between a {\it Hydra} run of 1 million particles
and {\it Enzo (PPM)} for different grid sizes. It is clear that a grid
size of $20^3$ produces significantly poorer results than the {\it
Hydra} data whereas a grid size of $100^3$ produces significantly
better results particularly in the low density region.  Although the
situation is confused by the lack of points available with {\it Enzo},
a grid size of $20^3$ can produce no more than 20 distinct values
across the length of the volume being modelled, it is clear that this
is too few to recover this model. However, even with $50^3$ cells in
the box the major features are largely recovered.  To make further
progress we require a more quantitative way of comparing the results
from the codes. To achieve this, we employ a cubic spline to
interpolate the data from all runs to the same 178 $x$ points at which
we have calculated the analytical values. The residue between each new
curve and the analytical solution was then summed and divided by the
number of points. We considered the residue both across the whole
region of interest from [-0.15, 0.15] and just across the shock-front
from [-0.13, -0.06].

\begin{table}
\caption{Residue of from the analytical solution of the Sod shock density in the planar [1,0,0] set-up for different static grid sizes and SPH resolutions for {\it Enzo, Hydra} and {\it Gadget2}. }
\begin{tabular}{ccc}
& {\it Enzo (PPM)} residue & {\it Enzo (Zeus)} residue\\
Grid size & Interface (Shock) & Interface (Shock) \\
\hline
$10^3$ & 0.343 (0.383) &  0.238 (0.220)\\
$20^3$ & 0.232 (0.166) &  0.173 (0.198)\\
$30^3$ & 0.164 (0.164) &  0.165 (0.172)\\
$40^3$ & 0.129 (0.125) &  0.166 (0.174)\\
$50^3$ & 0.095 (0.089) &  0.119 (0.130)\\
$60^3$ & 0.094 (0.110) &  0.114 (0.118)\\
$70^3$ & 0.076 (0.082) &  0.105 (0.126)\\
$80^3$ & 0.086 (0.082) &  0.081 (0.090)\\
$90^3$ & 0.061 (0.075) &  0.088 (0.088)\\
$100^3$ & 0.069 (0.064) & 0.085 (0.079)\\
AMR run ($400^3$) & 0.032 (0.030) & 0.035 (0.036))\\
\hline
\end{tabular}
\begin{tabular}{ccc}
& {\it Hydra} residue & {\it Gadget2} residue\\
No. particles & Interface (Shock) & Interface (Shock) \\
\hline
8M $(234^3,148^3)$ &  0.0559 (0.065) & \\
1M $(117^3,74^3)$  &  0.0852 (0.104) & 0.0817 (0.089)\\
250k $(74^3,48^3)$ & 0.112 (0.121) & 0.113 (0.102)
\end{tabular}
\label{table:sod}
\end{table}

The top part of table~\ref{table:sod} shows the residues from {\it
Enzo} over these ranges for increasing static grid size. As
Figure~\ref{fig:sod_comparison} showed, the overall fit across the
whole region of interest improves rapidly with the number of grid points.
The improvement for the AMR run highlights one of the greatest assets
of adaptive grid coding; the ability to tag cells for refinement based
on slopes or shock-fronts, allowing the extra resolution to be
concentrated on these problematic areas.  Table~\ref{table:sod} also
shows that {\it Enzo (Zeus)} typically has poorer results than {\it
Enzo (PPM)} for equal numbers of grid cells, something that came out
in Figures~\ref{fig:sod_density} and \ref{fig:sod_all}, and is a
result of energy loss due to the inaccurate treatment of the shock
jump conditions by {\it Enzo (Zeus)}.

The lower part of table~\ref{table:sod} shows the residues for {\it
Hydra} and {\it Gadget2} at different resolutions. Both SPH codes show
similar results, with similar residues in each at the same resolution.
If we compare these values to the nearest {\it Enzo} result we can get
an estimate for the number of particles needed per grid cell in the low density region for
similar recovery of the shock profiles. Included after the number of
particles for each of these runs is the effective number of particles
in the high and low density regions. As is obvious from this table and
Figure~\ref{fig:sod_comparison} roughly matching accuracy occurs
somewhere between these two limits. This is equivalent to when there
is approximately one SPH particle per grid cell.

The analysis done in this section does have one obvious flaw;
introducing adaptive meshes allows the grid-based codes to achieve a
high resolution while employing many less cells overall. Our
equivalence of one SPH particle per cell applies to cells in the high
resolution region not the total number of cells in the simulation. As such, this comparison works best in situations where the resolution is needed in the highest density areas. In
both this problem type and the one described in the next section, the
high density areas are not those that require higher resolution,
rather the region of interest is that where the density is changing
rapidly. Since usually SPH particles are tied to the mass flow of a system,
they are not able to increase the resolution over shock fronts, unlike
their grid counterparts. However, in many astrophysical problems, such
as those we shall meet in section~\ref{sec:cluster}, this is not
necessarily a serious limitation.

\subsection{The Sedov Blast Wave Test}

\begin{figure*}
\centering
\includegraphics[width=\textwidth]{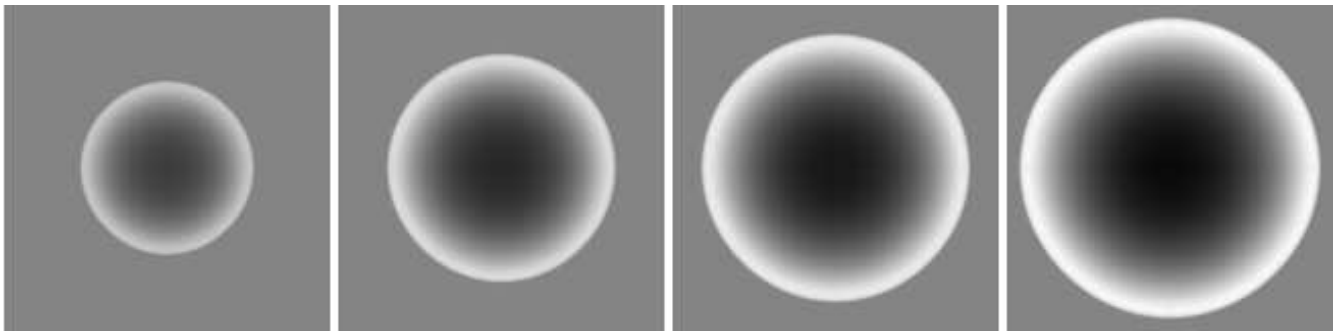}
\caption{Density projections of the Sedov blast test performed using
  {\it Enzo (PPM)} at $t=0.025, 0.05, 0.075$ and $0.1$.}
\label{fig:sedov_prop}
\end{figure*}

The Sedov Blast Test \citep{Sedov1959} is an intense explosion caused
by a quantity of energy deposited in the centre of the simulation
box. The result is a strong spherical shock that propagates through
the background homogenous medium.

This test is particularly appropriate in astrophysics since it
represents well the physics required to deal with supernova
explosions. An inability to resolve the resulting shock front will lead
to an incorrect estimate of the energy deposited into the galaxy and
the volume it affects.

It also poses different problems to both the particle and grid schemes
described above. Although the shock front expands as it travels
outwards it also sweeps up particles which increase the density
contrast of the shell relative to the ambient medium. The particle
codes will therefore find it progressively easier to reach the
required resolution. Since the shock front is also spherical, grid
codes will have to cope with any artificial grid-alignment effects. It
is therefore a test that is both appropriate and taxing.

The analytical calculation of the shock propagation is described in
full in \citet{Sedov1959} and \citet{Landau1959} who show the shock
front's radius is given by:

\begin{equation}
\label{eq1}
r(t) = \left(\frac{E_0}{\alpha\rho_0}\right)^{1/5}t^{2/5}
\end{equation}

\noindent where $E_0$ is the initial energy injected, $\rho_0$ is the
background density and $\alpha = 0.49$ for an ideal gas with $\gamma =
5/3$.

Since there is no cooling or gravity in this test, the quantities
above are all unitless. In our calculations, the initial density of
the medium is $\rho_0 = 1$, the explosion energy $E_0 = 10^5$. The
shock expands into a box of side $10$ and we compare our results at
$t=0.1$.  Figure~\ref{fig:sedov_prop} shows density projections (density integrated along the length of the simulation box) of the
simulation box during the Sedov test for times from $0.025$ to $0.1$.

\subsubsection{General Results}

\begin{figure*}
\centering
\includegraphics[angle = 270, width=\textwidth]{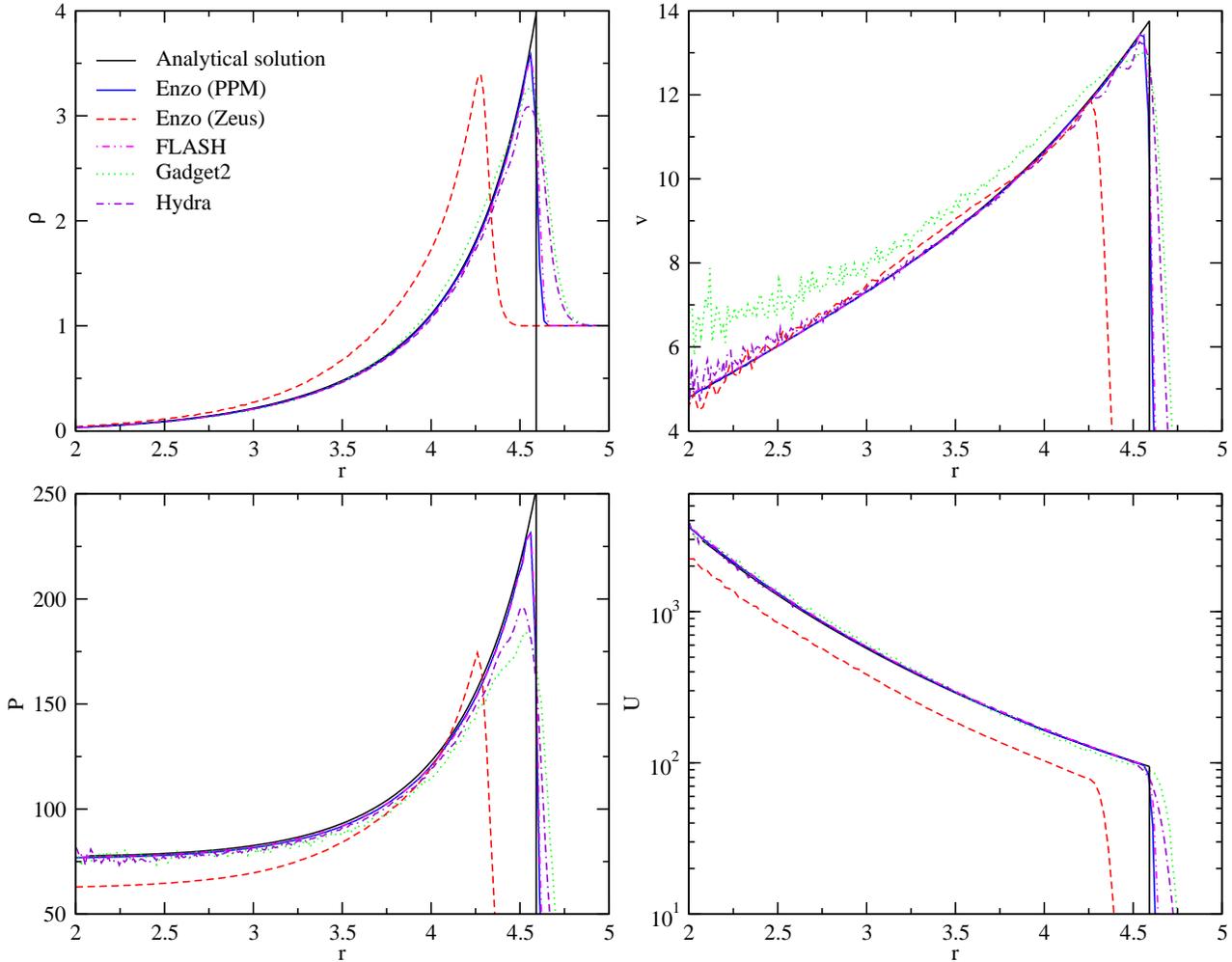}
\caption{Results from the Sedov blast test for each code at $t=0.1$. Clockwise from top left shows density, velocity, pressure and internal energy. The black solid line marks the analytical solution.}
\label{fig:sedov_plots}
\end{figure*}

\begin{figure*}
\centering
\includegraphics[width=\textwidth]{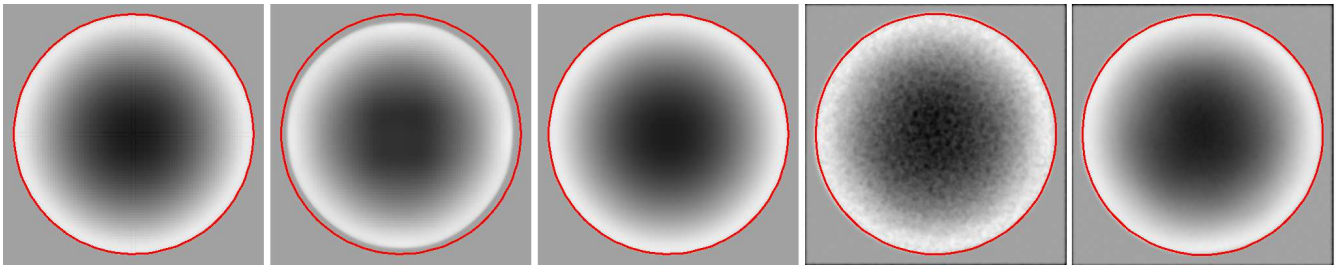}
\caption{Density projections of the Sedov blast test at $t = 0.1$ for
  all codes, with the expected shock position marked by the red circle. Left to right shows {\it Enzo (PPM)}, {\it Enzo (Zeus)},
  {\it Flash},
  {\it Gadget2} and {\it Hydra}. The density ranges from [$10^{0.6}$, $10^{1.23}$].}
\label{fig:sedov_projs}
\end{figure*}

Figure~\ref{fig:sedov_plots} shows the results from the Sedov blast
test at $t=0.1$ for each code and the analytical solution. For this test,
{\it Enzo} and {\it Flash} used an initial grid of $100^3$ with two levels of refinement,
each reducing the cell size by a factor of two. These subgrids were
placed where the shock front was detected. This gave a minimum cell
size of $0.025$ over the shock. For the AMR codes the central energy
was injected into a spherical region of radius 3.5 grid cells on the
finest levels, corresponding to $r=0.0875$, a common choice for this test.  Both {\it
Hydra} and {\it Gadget2} used 1 million particles relaxed into a
glass, with the energy added to the central 32 particles in a top hat
distribution.

We have deliberately chosen a large energy jump for our Sedov blast in
order to make this test challenging. A short discussion is added to
the end of this section describing some of the pit-falls we
encountered while attempting to successfully complete it. What is
immediately apparent from Figure~\ref{fig:sedov_plots} is that {\it
  Enzo (Zeus)} performs significantly worse on this test than all the
other codes, with its shock edge lagging substantially behind the analytical position. It shows a significant
energy loss in the first few timesteps, loosing 25\%
of its internal energy during this stage. In the previous section, we saw
that {\it Enzo (Zeus)} spread the shock front out slightly more than
{\it Enzo (PPM)} and {\it Flash}, which both utilise Godunov schemes,
but the difference was minimal and the shock-front was still sharp and
well represented. Here, however, we see that {\it Enzo (Zeus)}
underestimates the shock position at $t=0.1$ by almost 4\%. The cause of this significant energy loss is the production of a diamond-shaped, rather than spherical, shock front at very early times. This is discussed further in section~\ref{sec:code_spec}. Plots of the internal energy and pressure in Figure~\ref{fig:sedov_plots} also add to illustrate this effect, showing the effect of {\it Enzo (Zeus)}'s early energy loss and the resulting low pressure prior to the shock front. Both {\it Enzo (PPM)} and {\it Flash} perform this test extremely well, matching
the analytical solution to the shock's edge. The two SPH codes successfully recover the
location of the peak of the Sedov blast shell but it is smoothed out
in the radial direction, producing a lower peak density and a broader
shell. For the {\it Gadget2} run the gas velocity interior to the blast appears to be enhanced in Figure~\ref{fig:sedov_plots}. This is an artifact of the
entropy scatter discussed below. The gas interior to the blast front will rapidly sort itself in entropy, resulting
in smooth density and temperature profiles, leaving the velocity profile disordered (and so enhanced).

Figure~\ref{fig:sedov_projs} shows density projections of the final state of the simulations at $t=0.1$. The analytical position of the shock front is shown by the solid red line. As can be seen on Figure~\ref{fig:sedov_projs} although the location
of the shock front in the {\it Gadget2} figure is well recovered there
seems to be a lot of ``noise'' producing a grainy appearance to the shaded image. Although energy and entropy are well
conserved by {\it Gadget2} for extreme shocks like this one the {\it
  Gadget2} viscosity implementation introduces a lot of entropy
scatter. The high entropy particles are too hot for their
surroundings and drive small bubbles, producing the structure seen in 
Figure~\ref{fig:sedov_projs}. This feature of {\it Gadget2} does not
appear for more normal shock jumps such as that studied in the
previous section or for less energetic Sedov blasts. We have tried
several approaches to resolve this problem which are mentioned in
section~\ref{sec:code_spec} below for those interested.

\subsubsection{Quantitative Comparison}

\begin{figure*}
\centering
\includegraphics[angle=270,width=\textwidth]{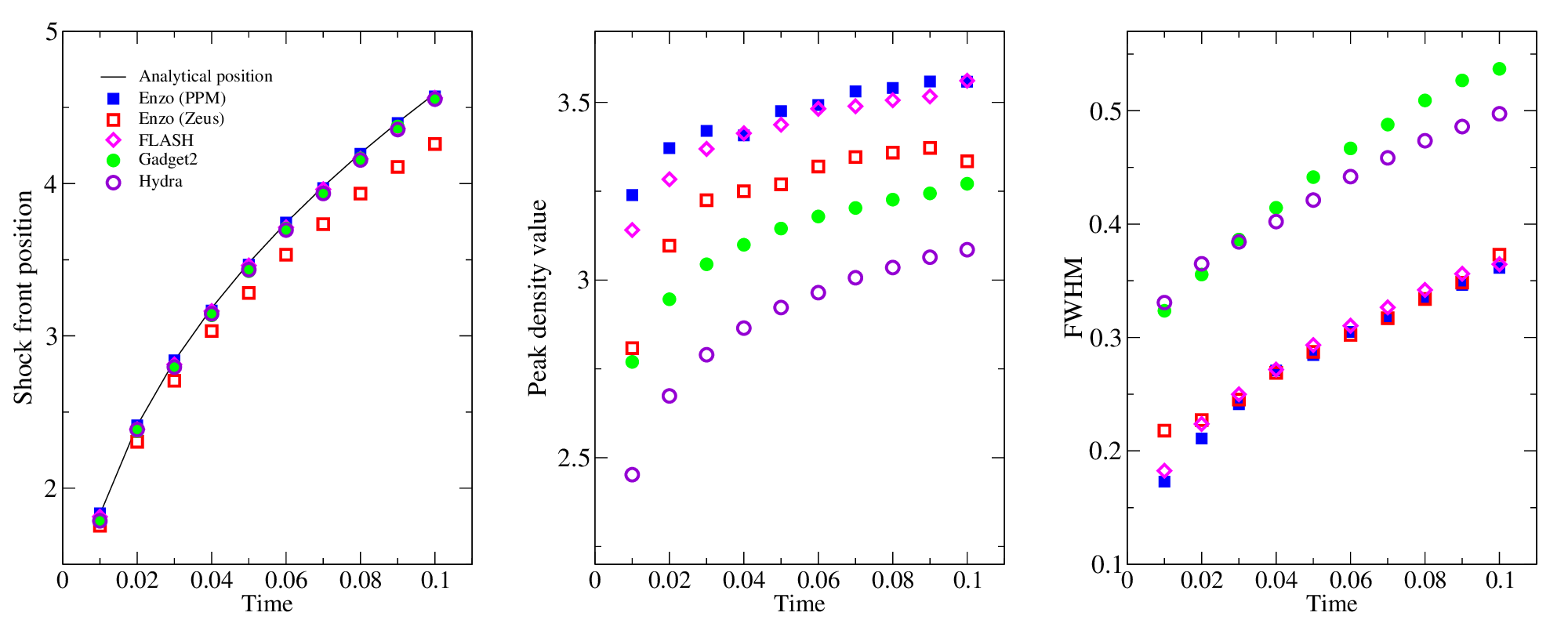}
\caption{For the Sedov blast test, from left to right; position of shock front over time, the maximum density value and the width of the shock front at half maximum.}
\label{fig:sedov_anyl}
\end{figure*}

As the previous section showed, four of our codes appear to do a reasonable job
of following the location of the the Sedov blast front. Due to its
inability to conserve energy initially, {\it Enzo (Zeus)} does not perform this test well and only achieves its current results by adapting code parameters (see ``code-specific issues''
below). In order to quantitatively compare the codes we have measured
the position of the peak of the density, the maximum density obtained
and the width of the shock front at half this maximum density as a
function of time. The results of this exercise are given in
Figure~\ref{fig:sedov_anyl}. We see that {\it Enzo (PPM), Flash, Hydra}
  and {\it Gadget2} are all capable of accurately following the location of
the shock front. As is visible on Figures~\ref{fig:sedov_plots} and \ref{fig:sedov_anyl} the two
SPH codes do not obtain the high maximum densities recovered by the
AMR codes. {\it Gadget2}'s maximum peak position is lower than {\it Enzo (PPM)} and {\it Flash} by 8\% and {\it Hydra} is 13\% less. {\it Enzo (Zeus)}'s lag in the shock front propogation makes its position equivalent to a time of $t=0.085$, at which time it is between the PPM-based schemes and the SPH techniques with a top density value 6\% below {\it Enzo (PPM)} and {\it Flash}.

The lower SPH peak density is not a surprise as the inherent smoothing of the SPH
method broadens the shock front and lowers the maximum recoverable
density. The AMR codes have the ability to insert additional
resolution elements where quantities are rapidly changing rather than where
lots of mass has piled up and this acts to improve their fit to the
analytic solution in the region of the blast. Despite this disadvantage, the SPH codes correctly reproduce the shock front position and, at around 10\%, the differences in their peak density values are remarkably small.

It is difficult to undertake a ``number of particles per grid cell''
comparison similar to the one above for this test for two reasons:
firstly, the initial conditions rapidly become impossible to set up in
a spherically symmetric way for the grid codes if they are not allowed
to refine heavily.  Our initial condition set-up inserts the energy
into a region of radius 3.5 grid cells (or 180 cells). If fewer than
this number are used the cells do not represent a spherical energy
input well. Without refinement a region encompassing this many cells
rapidly becomes large, whereupon the initial conditions no longer represent a Sedov explosion. In addition, small numbers of large cells
simply cannot be expected to follow a spherical blast and little
useful is learnt. Secondly, the SPH codes are helped at late times
because the Sedov blast ``sweeps up'' the particles it encounters (as
it is supposed to) which effectively increases the number of SPH
resolution elements in the shock front, as is to be expected given the
density enhancement. The AMR codes naturally account for this by
adding extra layers of refinement. However, if an unrefined comparison
is attempted this high level of refinement has to be present
everywhere and at all times, so an incredible number of cells are
needed. For this particular test we find that we need static grids of
order $250^3$ to reproduce a blast with similar resolution to an SPH
model with only $100^3$ particles.  The quantity that should be
directly compared is the number of cells per particle in the region of
interest. As above we find that roughly one SPH particle per AMR cell
is required to obtain similar effective resolutions.

\subsubsection{Code-Specific Issues for the Sedov blast}
\label{sec:code_spec}

\begin{figure}
\centering
\includegraphics[width=\columnwidth]{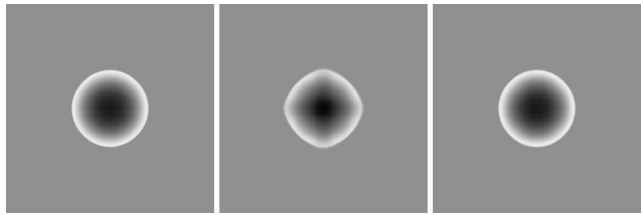}
\caption{Density projection showing the early evolution of the Sedov blast test at t = 0.01 for the AMR codes. Left to right shows {\it Enzo (PPM)}, {\it Enzo (Zeus)} and {\it Flash}. The projected density range is [$10^{0.87}, 10^{1.1}$]. At this time, the assymmetry of the shock-front with {\it Enzo (Zeus)} is clearly visible. Both SPH codes give spherically symmetric results.}
\label{fig:zeus_diamond}
\end{figure}

Figure~\ref{fig:zeus_diamond} shows the density projection of the Sedov blast test performed by {\it Enzo (Zeus)} at early times. The problem becomes immediately apparent; instead of being spherical, the shock-front is an assymetrical diamond shape. Over time the shock-front becomes spherical, as can be seen in Figure~\ref{fig:sedov_projs}, and the dramatic energy loss stops, but by this stage 25\% of the shock's energy has been lost. The reason {\it Enzo (Zeus)} fails to produce a spherical shock-front at early times is not obvious. It appears overly sensitive to the initial set-up not being a perfect Sedov blast start (since a point-like energy injection can only be approximate). Many different solutions were attempted, including injecting the energy as a Gaussian rather than `Top hat' profile and adding it into a larger radius (which helped, but did not greatly improve the situation). Even to recover the pretty miserable results that {\it Enzo (Zeus)}
achieved above we had to use a somewhat dirty fix of increasing the code's quadratic artificial viscosity term, $QV$. This value is a broadening parameter that controls how many cells the shock is spread over. By default $QV=2$, which is used for the tests in sections~\ref{sec:sod} and \ref{sec:cluster}. \citet{Agertz2007} found that varying $QV$ made
little difference to the evolution of the fluid, but just broadened
the shock. In this case, we find that increasing the width of the
shock decreases the assymmetry at early times, although Figure~\ref{fig:zeus_diamond} shows the effect was far from perfect. Using the default value of $QV=2$,
the shock-front lags even further behind the result shown in
Figure~\ref{fig:sedov_plots}, which uses $QV=10$. The increase in $QV$
comes at a price; the spreading of the shock over more cells weakens
it, causing the peak density to be lowered. Raising $QV$ beyond 10
improves the position of the shock further, but the value now becomes
unphysically high and we do not recommend using it.

The Sedov blast has already been used by \citet{Springel2002} to test
{\it Gadget2}, and indeed for the parameters \citet{Springel2002} used
{\it Gadget2} works beautifully and does not exhibit any spurious entropy
driven bubbles. In order to investigate their origin we tried
injecting the energy into a single point particle or smoothing the
injection profile using a Gaussian function (with a cut-off at a few percent
of its central value) rather than a top hat, and also changing the number of 
SPH neighbours in order to help numerical convergency. None of
these changes makes any practical difference to our test and the
entropy driven bubbles still occurred. In all the Gadget2 runs performed for this test, the artificial viscosity parameter ArtBulkVisc has been set to the frequently used value of 1; there are hints that a choice for a higher value might lead to a sensible reduction of the entropy driven bubbles (see Pakmor et al., in preparation). It appears that in extreme
shocks the viscosity implementation employed by {\it Gadget2} does not
sufficiently prevent particle inter-penetration and we warn users
of {\it Gadget2} to be wary of the generation of spurious entropy
scatter in the vicinity of extreme shocks.

\section{Gravitational Tests}
\label{sec:cluster}

In this section, we move away from the formation and resolution of
shocks to look at a new aspect of the codes; how they deal with
gravity. While treatment of fluids is important, few astrophysical
simulations can be performed without a self-gravitating gas. However,
adding self-gravity, where every fluid element is affected by every
other, dramatically complicates the situation and it is not possible
to design a test with an exact analytical solution anymore. Since it
is still essential for the purpose of this comparison that our
problems remain well-posed, we select situations in which the correct
behaviour of the system is known, even if it cannot be mathematically
expressed. To do this, we perform two tests; the first of these
concerns a static gas profile in equilibrium. Gravity acts to
try and collapse the gas, while pressure pushes it outwards. While
these forces remain perfectly balanced, the gas remains at rest. This
situation is analogous to a relaxed galaxy cluster and requires the code to
resolve the gas density over many orders of magnitude. The second test involves
the same cluster translating through the box. By using periodic boundary conditions, the cluster's velocity 
is chosen such that it should return to its original position after 1 Gyr. With no external forces,
the cluster should remain in hydrostatic equilibrium and retained its profile during the translation.

\subsection{Initial conditions for the cluster}

The model used for the galaxy cluster is the King model
\citep{King1966, Padmanabhan2002}, which was chosen because it possesses a
finite radial cut-off, and is therefore a well defined problem for a
code comparison. Its form is based on the distribution function:

\begin{equation}
\label{eq2}
f(\epsilon) =  \left\{ \begin{array}{ll}
  \rho_c (2\pi\sigma^2)^{-3/2}\left(e^{\epsilon/\sigma^2}-1\right) & \epsilon \ge 0,\\
  0 & \epsilon < 0
  \end{array} \right.
\end{equation}

\noindent where $\epsilon = \Psi - \frac{1}{2}v^2$, is the coordinate
change for the shifted energy, $\rho_c$ is the central density and
$\sigma$ is related to (but not equal to) the velocity dispersion. The
resulting density distribution of this cluster vanishes at the
\emph{tidal radius}, $r_t$. Integrating over all velocities yields a
density distribution:

\begin{equation}
\label{eq3}
\rho\left(\Psi\right) = \rho_c\left[e^\frac{\Psi}{\sigma^2}{\rm erf}\left(\sqrt{\frac{\Psi}{\sigma^2}}\right)-\sqrt{\frac{4\Psi}{\pi\sigma^2}}\left(1+\frac{2\Psi}{3\sigma^2}\right)\right].
\end{equation}

\noindent Putting this into the Poisson equation results in a second
order ODE which can be solved numerically.  This model has three independent parameters, the
mass of the cluster, the tidal radius and the concentration $c=
\log_{10}(r_t/r_0)$, where $r_0$ is the central or \emph{King
radius}. For this problem, we selected a concentration of 3, $r_t =
1$\,Mpc and a cluster mass of $10^{14}$\,M$_\odot$. This results in a
King radius of $1$\,kpc. Therefore, to successfully
maintain hydrostatic equilibrium, the codes must be able to model the
cluster out to $1$\,Mpc while resolving the $1$\,kpc core. This makes
it a particularly challenging test.

\subsection{The Static Cluster}

\begin{figure*}
\centering
\includegraphics[angle=270,width=\textwidth]{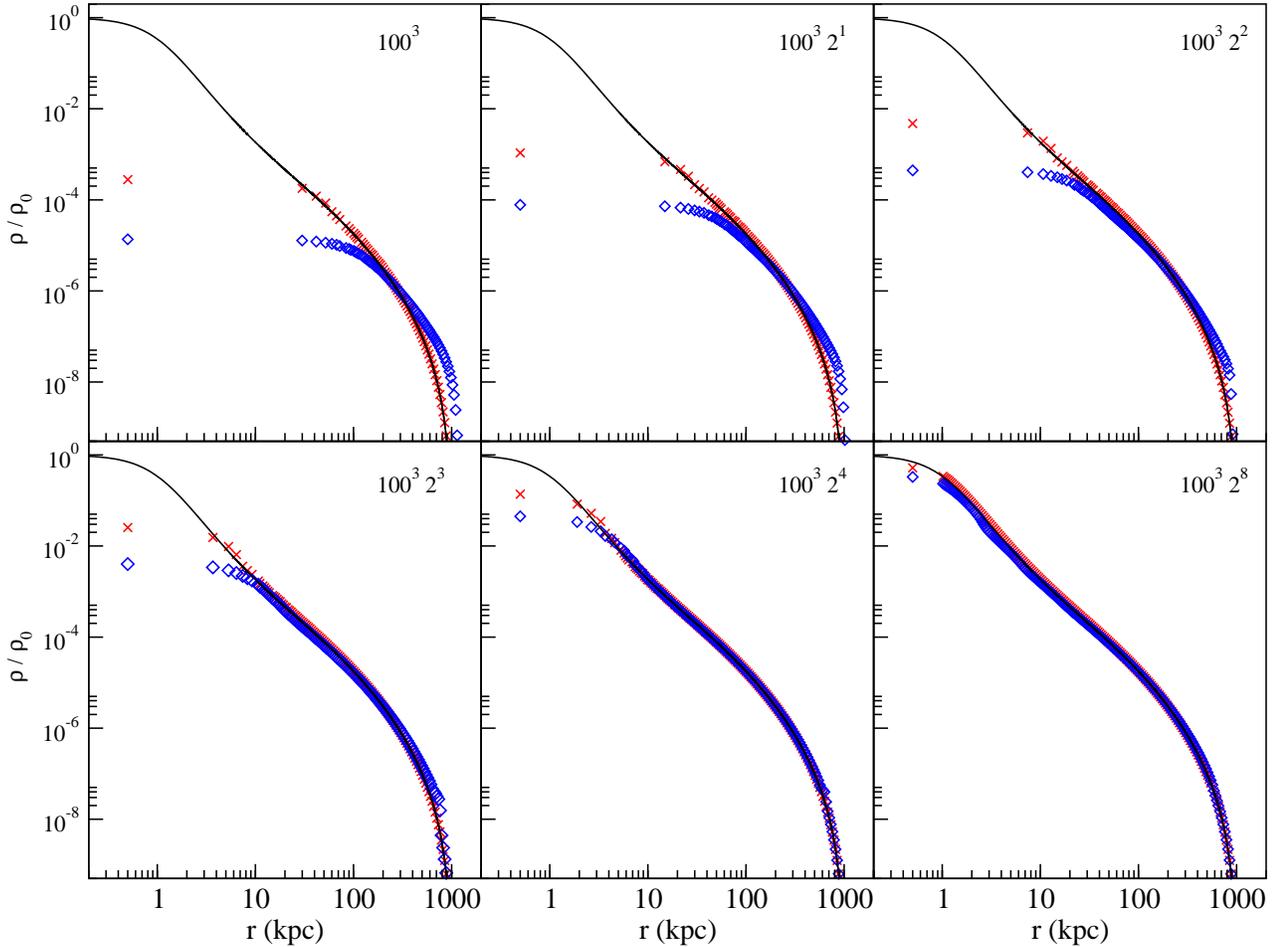}
\caption{Density profiles of the static King-model cluster run with
{\it Enzo (Zeus)} with increasing resolution. The solid line shows the 1D calculated 
numerical solution for the profile, with the red crosses and open blue boxes indicating the simulated profile at 0\,Gyr and 1\,Gyr, respectively. The number to the top right of
each panel indicates the size of the root mesh and the number of
refinements (i.e. the bottom right panel has eight levels of refinement
each increasing in resolution by a factor of 2).}
\label{fig:king_res}
\end{figure*}

\subsubsection{Resolution of the cluster}

The two key requirements for success in this test are to be able to
resolve the core and to have an accurate gravitational solver.
Figure~\ref{fig:king_res} shows this case in point. It plots the
density profile of the cluster at the start of the simulation and
after 1\,Gyr for steadily increasing levels of resolution. Although
the King model does not have an analytical solution, a one-dimensional
numerical solution for the cluster's profile can be achieved from a
simple numerical integration. This is shown in each plot as the solid
line. In the top left image, the cluster is modelled on a static grid
of size $100^3$. The initial match to the profile is good up to
densities of $10^{-4}$, but the inner kiloparsec that
contains the core is totally unresolved. The effect is for the
pressure to dominate over the gravity and the cluster expands,
dropping the density in the core still further. After 1\,Gyr, the
cluster's profile is barely recognisable, despite the lack of external
forces. Moving one plot to the right in Figure~\ref{fig:king_res}, we
see the results of adding a subgrid into the area that contains the
highest density, i.e. the centre of the cluster. The density in the
central region is now followed up to 0.15\,M$_\odot$pc$^{-3}$. The
cluster is still not balanced and deviates away from the profile, but
the shift is markedly less. As we continue to add in levels of
refinement, we see the central density rise to meet the numerical
expectation as the core becomes more resolved. The profile of the
cluster changes progressively less until we reach five levels of
refinement, when the change becomes almost undetectable except at the
very centre.

\subsubsection{General results}

\begin{figure*}
\centering
\includegraphics[angle=270,width=\textwidth]{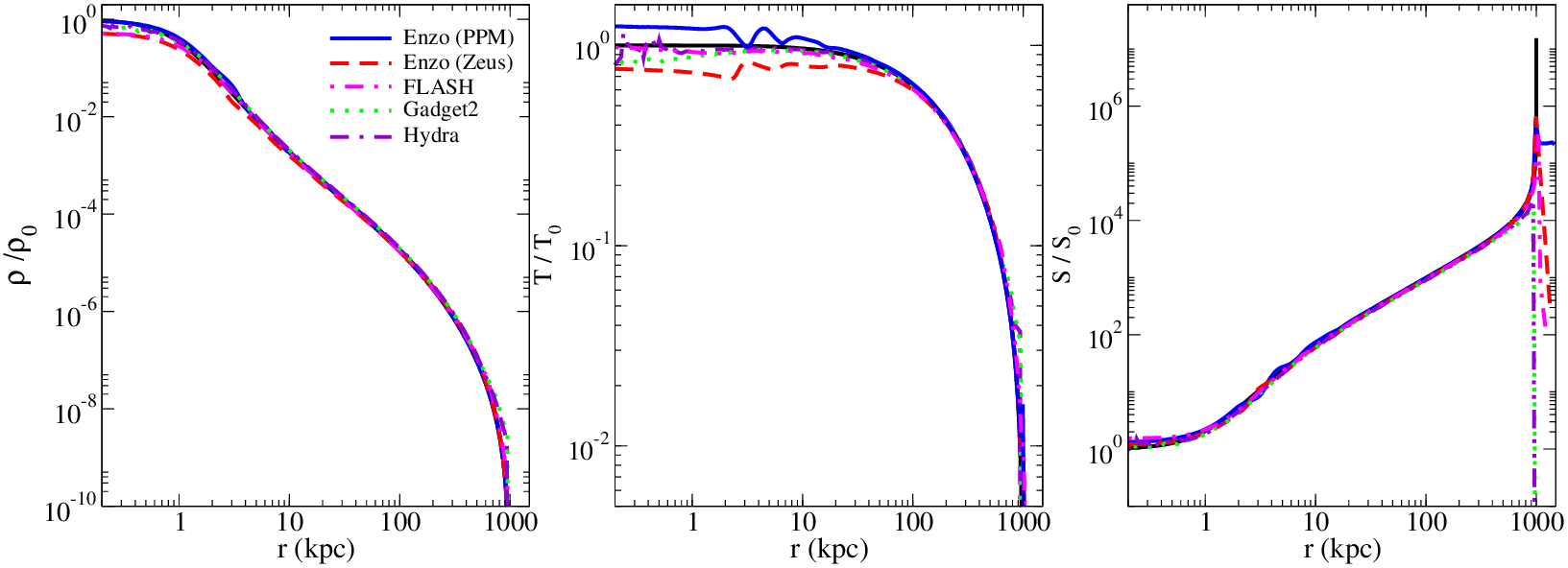}
\caption{Profiles from the static King-model cluster after
1\,Gyr. Left to right shows density, temperature and entropy. The black line shows the expected profile. In this test, {\it Enzo} used an inital grid of $100^3$ and 8 levels of refinement (minimum cell size, $\Delta x_{\rm min}$, of 0.12\,kpc), {\it Flash} used an initial grid of $128^3$ with 7 levels of refinement ($\Delta x_{\rm min} = 0.18$\,kpc) and {\it Gadget2} and {\it Hydra} used 100,000 particles. }
\label{fig:static_king}
\end{figure*}

Figure~\ref{fig:static_king} shows the density, temperature and
entropy profiles for the cluster after 1\,Gyr. For this test, {\it Enzo}
was run with an initial grid of $100^3$ and eight additional levels of refinement,
each reducing the cell size by a factor of two. These subgrids were
placed anywhere where the cell mass was above a critical value. The
cluster was set up in a box of size 3\,Mpc with isolated
gravitational boundary conditions. This gave a minimum cell size
in the core of 0.12\,kpc. {\it Flash} ran with the same boxsize and with periodic boundary conditions. It used a slightly larger initial grid of $128^3$  (since the gravity solver requires factors of two) and included seven additional levels of refinement, each of which reduced the cell size by a factor of two. This gave a minimum cell size in the core of 0.18\,kpc. {\it Hydra} and {\it Gadget2} radially
perturbed a glass of 100,000 particles to the desired density profile
within a periodic 3\,Mpc box.

\begin{figure*}
\centering
\includegraphics[width=\textwidth]{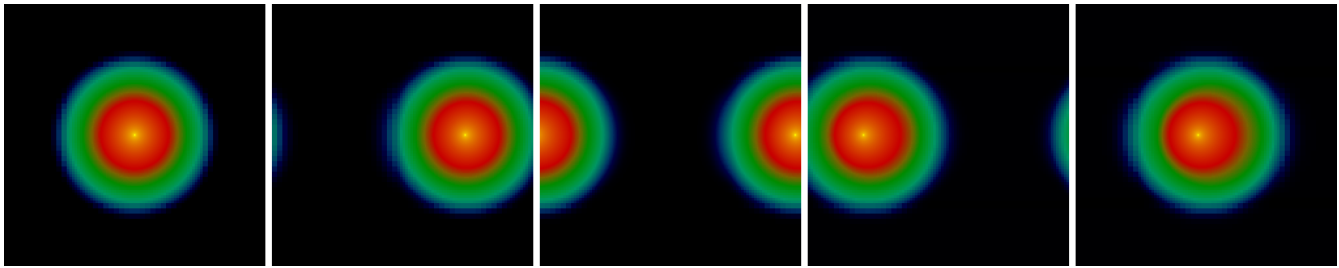}
\caption{Density projections of the cluster over the course of 1\,Gyr in which is moves once around the simulation box. Images taken at 0, 250, 500, 750, 1000 Myrs with projected density range [$10^{8.4}$, $10^{16.5}$]\,M$_\odot$ Mpc$^{-2}$. Yellow and red shows higher density regions than green while black is very low density. (Images produced with {\it Enzo (Zeus)}).}
\label{fig:king_moving_proj}
\end{figure*}

\begin{figure*}
\centering
\includegraphics[width=\textwidth]{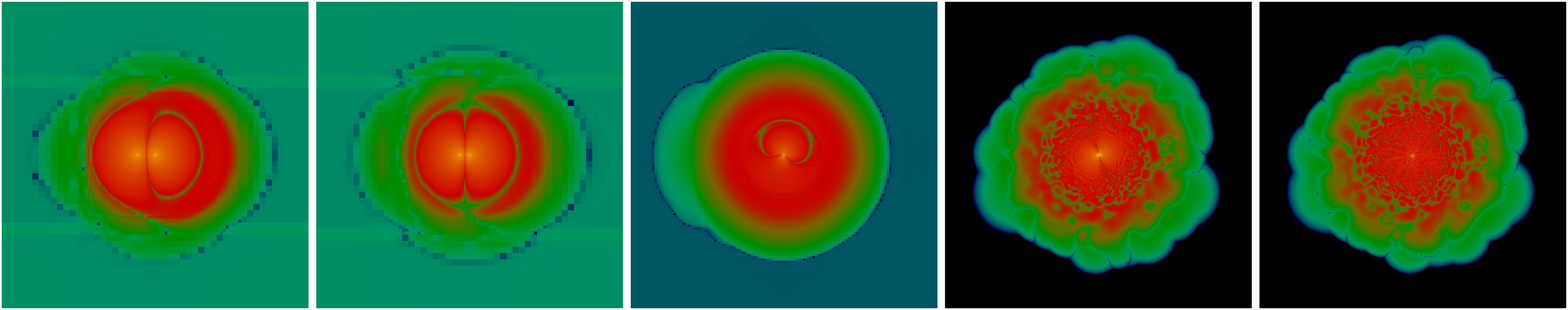}
\caption{Image subtractions of the density projections at the start
  and end of the translating cluster test. From left to right shows
  {\it Enzo (PPM)}, {\it Enzo (Zeus)}, {\it Flash}, {\it Gadget2} and {\it Hydra}. The projected density range is [$10^{4}$, $10^{18.6}$]\,M$_\odot$ Mpc$^{-2}$. }
\label{fig:king_moving_subtract}
\end{figure*}

As can be seen in the density profile in Figure~\ref{fig:static_king},
with sufficient central resolution all the codes are successful at
keeping the cluster in equilibrium and resolve the core well. All the codes show negligible deviations from the static King profile after 1\,Gyr, matching the analytical profile over seven orders of magnitude, down to densities of
$10^{-10}$.  At the low
density edge of the cluster, small deviations ($\sim 15\%$) from the profile are seen
as the cluster edge diffuses into the background. {\it Enzo (PPM)},
{\it Enzo (Zeus)} and {\it Flash} produce the closest fit to the numerical solution at these densities, owing
to the larger choice of initial grid which fixes the minimum
resolution. {\it Hydra} and {\it Gadget2} show small deviations close to the cluster's tidal radius as they start to run out of particles. 

Unlike the shock tests in section~\ref{sec:shock}, all our codes in this test and in section~\ref{sec:translatingking} refine based on increasing mass. Although grid based codes
can make resolution choices from other parameters, mass in a grid cell (i.e. density) is the most
used for simulations on scales above a few parsecs.  However, in
contrast to the particle based codes, a grid code's minimum resolution
is always known, since it is fixed by the specified minimum grid
size. Depending on the simulation type, a fixed minimum resolution is either a huge advantage or a drain on computational time. In simulations where the denest structures are the primary focus, SPH's
ability to sweep all the particles into these areas means that high
resolution can be gained quickly and efficiently. To gain the same
efficiency from a grid code, a careful choice of minimum resolution grid and
refinement criteria must be made. Both the test in this section and the following one are examples of this problem type and were significantly more time consuming for grid based codes than for particle based ones. This is in contrast to the shock tests which ran
faster. However, if the focus of a simulation is the surrounding structure of a dense object, then the ability to specify a minimum resolution is extremely useful and time efficient. In a particle scheme this would normally be obtained by adding enough additional particles such that both the dense and surrounding medium were resolved; a solution which may significantly add to the expense of the simulation.

Figure~\ref{fig:static_king} demonstrates the two SPH codes
achieve a similar result with 100,000 ($46^3$) particles to that
obtained by the AMR codes with initial meshes of around 1 million cells
and 8 or 9 levels of refinement. To achieve this with a static mesh
would be impossible, as the mesh would need to be $25,600^3$. Not
surprisingly then, given the complication of placing and interpolating
between refinements the SPH codes are much faster. We note that
in the SPH runs the distance from the centre to the 32nd particle is
0.4\, kpc, which is very close to twice the minimum cell size
of the AMR runs in the core. Again, we see that in order to obtain
convergence between the AMR and SPH codes we require roughly one
particle per cell in the region of interest. Unfortunately for the AMR
codes, for this particular problem which includes a wide density range
the ability of SPH codes to naturally increase resolution with mass is
 very advantageous.

Finally for this section, Figure~\ref{fig:static_king} also clearly
shows that both of the {\it Enzo} implementations include noticeable
ringing in the temperature profile which is reflected in the entropy
and less obviously density profiles. This causes a deviation from the analytical solution of around $\pm 20$\% in the inner 2\,kpc (10 cells). This ringing is not present in the
analytic solution and doesn't appear in either {\it Flash} or the SPH
codes which closely resemble both each other and the calculated
solution. 

\subsection{The Translating Cluster}
\label{sec:translatingking}

\begin{figure*}
\centering
\includegraphics[angle=270,width=\textwidth]{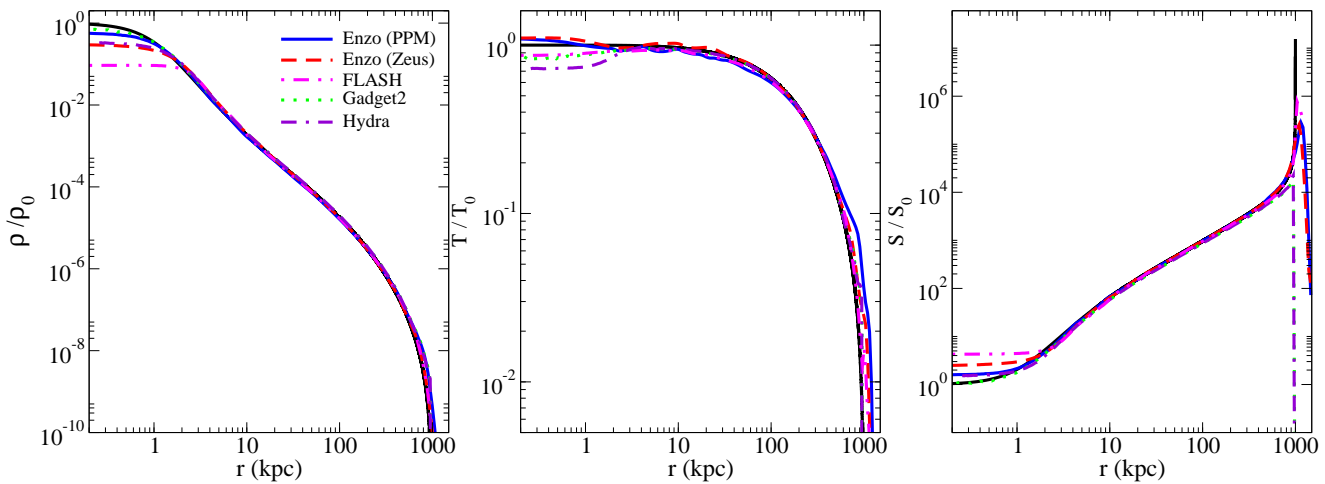}
\caption{Profiles from the translating King-model cluster after
1\,Gyr. Left to right shows density, temperature and entropy. The black line shows the expected profile. In this test, {\it Enzo} used an initial grid of $50^3$ and 9 levels of refinement (minimum cell size, $\Delta x_{\rm min}$, of 0.12\,kpc), {\it Flash} used an initial grid of $128^3$ with 7 levels of refinement ($\Delta x_{\rm min} = 0.18$\,kpc) and {\it Gadget2} and {\it Hydra} used 100,000 particles. }
\label{fig:trans_king}
\end{figure*}

Using the stable clusters developed above we can test the Galilean
invariance of the codes by giving them a velocity relative to the
static simulation volume. This is a commonly encountered situation for
cosmological simulations where large objects often move at many
hundreds of kilometres per second relative to the background. At
sufficiently high velocities it is well known and straightforward to
demonstrate that mesh based codes are not Galilean invariant, whereas
particle based methods are \citep[e.g.][]{Wadsley2008}. What we wish to investigate is the size of
these effects for typical astrophysical objects moving at typical
astrophysical velocities.

For this test the cluster was given a bulk velocity such that, in
1\,Gyr, it moved around the simulation box once, returning to its
original starting position. This is shown visually in
Figure~\ref{fig:king_moving_proj} which also demonstrates the box's boundary conditions which are periodic. Since there are no external forces
acting on the cluster, the end profile should be identical to the
initial one. 

The time consuming nature of this test for grid codes also became an issue. {\it Enzo (Zeus)} and {\it Enzo (PPM)} lowered their initial grid to $50^3$, but added an additional level of refinement maintaining a minimum cell size of $0.12$\,kpc. {\it Flash} managed, after a long run, to maintain the same resolution as for the static test, with a minimum cell size of 0.18\,kpc. The two SPH codes also had the same set-up as for the static case. 

Figure~\ref{fig:king_moving_subtract} shows the residue
remaining if the initial configuration is subtracted from the final
one. Another fundamental difference between grid based and particle based techniques is visible here in the background medium. Unlike a particle code, grid codes cannot have zero density and energy cells. Therefore, a cool, low density gas surrounds the cluster in
this test when run by both {\it Enzo} and {\it Flash}. In the case of Figure~\ref{fig:king_moving_proj}, which only contains results from {\it Enzo (Zeus)}, the minimum value shown is the background density, which is therefore displayed in black. From looking at the cluster core in Figure~\ref{fig:king_moving_subtract}, clearly neither of the {\it Enzo} simulations returns to the
correct position, with a particularly significant lag in the case of
{\it Enzo (PPM)}. In both these cases, the cluster as a whole correctly returns to the centre of the simulation box, but the central core does not, leaving it off-set from the centre of the cluster. This can be seen clearly in the last image of Figure~\ref{fig:king_moving_proj}. After 1\,Gyr, {\it Enzo (PPM)} has a core 176\,kpc from the original start position while {\it Enzo (Zeus)} finishes with its core 93\,kpc short of the initial position. The outer regions of the cluster also show distortion due to the lower resolution of the initial grid. These problems don't appear to occur for either the
SPH codes or {\it Flash}. {\it Flash} shows a small amount of distortion in the outer region of the cluster, although the central region stays largely intact with the core only 1.7\,kpc from the centre of the box. Both SPH codes show very similar results with {\it Hydra} getting its core closest to the original start position with a $0.4$\,kpc off-set and {Gadget2} with a $25$\,kpc off-set. In lower density regions, we see some distortion to the cluster's structure which becomes more marked at higher radii. This is due to the number of particles significantly decreasing as we move away from the core. 

Figure~\ref{fig:trans_king} is equivalent to Figure~\ref{fig:static_king} except that now the cluster has moved
once around the periodic box. All codes maintain the density profile of the cluster extremely well, with the only deviations appearing in the core. Here, {\it Gadget2} performs best, resolving the core in good agreement with the analytical prediction. All other codes struggle to resolve the core after 1\,Gyr, with {\it Flash} struggling the most and underestimating the core density by a factor of 10. Lower resolution runs (minimum cell size of 0.23\,kpc) using {\it Enzo} also showed a similar drop in core density in this test, whereas a much smaller difference was noticed for the static cluster at the same resolution. This change in core density was seen, but not understood, in the Santa Barbara Comparison Project \citep{Frenk1999} which observed the lower cluster densities in AMR codes. Without an analytical prediction for what the correct density should be, it was impossible to determine which of the recorded densities was more accurate. This test highlights two important effects controlling these core properties; resolution and advection. In the first instance, it is important to resolve the cluster core otherwise its size will be overestimated, leading to a lower central density. This is demonstrated clearest in Figure~\ref{fig:king_res} where in the case with four levels of refinement, the cluster remained intact but the central density still dropped. In the second case, the explicit movement of the fluid through the grid (i.e. advection) in AMR can (but not necessarily does) cause a drop in the central density. This is an explanation for why the static and translating king models differ in the AMR case, even at the same resolution. We see from Figure~\ref{fig:trans_king} that {\it Flash} suffers most from advection issues whereas {\it Enzo (PPM)} shows minimal difference from the static model, following the analytic profiles at least as well as the SPH codes. In both these areas, SPH's ability to follow the mass allows it to excel quickly and efficiently compared with AMR codes who have to be more careful when selecting their refinement criteria. 

While this test provides clues towards the deviation in cluster properties in the Santa Barbara Comparison Project, it should be noted that other effects can also be at work. In cosmological models, for instance, turbulence plays an important role as investigated by \citet{Wadsley2008} and \citet{Mitchell2008}. 

The temperature plot in the second panel of Figure~\ref{fig:static_king} shows that {\it Flash}, {\it Gadget2} and {\it Hydra} all underestimate the core temperature whereas {\it Enzo (PPM)} and {\it Enzo (Zeus)} overestimate it slightly. Additionally, all the AMR code overestimate the cluster temperature at its edge with {\it Enzo (PPM)} having the poorest fit. The core's entropy is also overestimated in the case of the {\it Flash} and {\it Enzo (Zeus)}; unsurprising since this depends on the inverse of the density. 

\subsubsection{Gravitational tests: Code-Specific Issues}
\label{sec:king_code_spec}

\begin{figure}
\centering
\includegraphics[width=6cm]{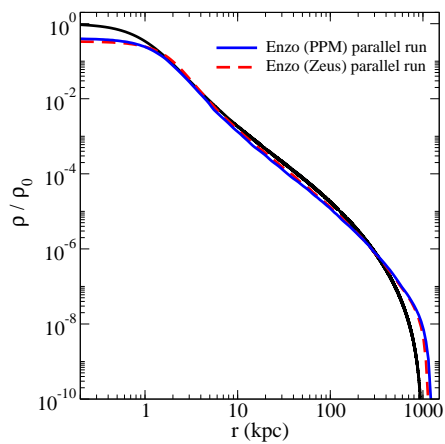}
\caption{Density profile from the translating king model after 1 Gyr run with {\it Enzo} in parallel. Run on multiple processors, the spurious velocity motions cause the gas to become over-mixed and the cluster density to drop in the core and expand at the cluster edge. The black line shows the expected profile.}
\label{fig:king_enzo}
\end{figure}

For both the static and translating cluster tests presented above, {\it Enzo} was run on a single processor. When using the parallised version of {\it Enzo}, the gravity solver suffers from excess noise in the velocity motions which mix the gas, reducing the entropy in the cluster centre. The result is a density drop in the cluster core and an expansion of its outer regions, unrelated to the issues discussed in section~\ref{sec:translatingking}. The resulting density profiles using the parallised version of {\it Enzo} are shown in Figure~\ref{fig:king_enzo}. In other situations  \citep[e.g.][]{Tasker2007} the gravity solver has been tested successfully to give identical results in serial and parallel, but an improved, less noisy, gravity solver is currently being developed to solve this problem where is occurs. 

{\it Enzo (PPM)} also experienced problems dealing with the background medium. The sharp drop in conditions between the cluster edge and low density background gas can cause {\it Enzo
(PPM)} to artificially produce negative densities and energies. This
was corrected for by setting a minimum density of
$0.1$\,M$_\odot$Mpc$^{-3}$, which negative densities were
automatically set to. Having to approximate densities multiple times during the run is a likely reason for the momentum loss that is clearly observed in Figure~\ref{fig:trans_king}.

The default multigrid algorithm used in the publicly available version
of {\it Flash 2.5}, proved too slow to be used practically. In order to
overcome this we implemented a hybrid FFTW based multigrid gravity solver,
resulting in a vast reduction in the runtime of the simulations. The latest
version of {\it Flash 3.0} includes a more efficient multigrid algorithm and a nested
FFTW gravity solver is currently under development for release in the near
future (similar to the one used here).

For {\it Gadget2} we encountered a problem in that although the radial
profiles were very stable the cluster as a whole tended to drift
around over time. This is due to the difficulty of very accurately
determining the lowest order term in the gravitational force expansion
for a treecode. Each individual term includes a small error, with
these errors largely but not exactly uncorrelated. Under these
conditions the total momentum of the system is not guaranteed to be
conserved exactly and a ``random walk'' occurs. As the configuration
is designed to be entirely static the direction of the residual force
is highly correlated from one step to the next and despite recovering
the correct value to one part in $10^8$ this still leads to a net
drift. It is possible to circumvent this, as shown above, by
dramatically reducing the opening angle for the tree but this rapidly
negates the advantage of using a tree in the first place. For more
normal simulations this tiny zeroth order force error is of course
negligible as the random motion of the particles disorders the
direction of the drift error as time progresses. $P^3M$ based gravity
solvers like that employed by {\it Hydra} do not suffer from this
problem as the zeroth order term in the Fourier transform is
automatically zero.

\section{Discussion and conclusion}

We have presented four well posed tests with known solutions which we
have used to compare four well used astrophysical codes; {\it Enzo,
  Flash, Gadget2} and {\it Hydra}. We have examined each code's
ability to resolve strong shocks and incorporate an accurate gravity
solver capable of resolving forces over many orders of magnitude in density. 

The tests presented here were specifically designed to be difficult and push the codes to their limit; appropriately so since many situations in astrophysics demand extreme physical conditions. Despite these requirements, all the codes did pass our tests satisfactorily and what we highlight here is the strengths and weaknesses of the different techniques.  

Significant issues were {\it Enzo (Zeus)}'s failure to produce a spherical shock, a problem that leads to a lack of energy conservation and to
poor recovery of the position of a Sedov blast. For extreme shocks the
artificial viscosity implementation of {\it Gadget2} leads to spurious
entropy driven bubbles and for very static configurations the
centre-of-mass can slowly drift due to the difficulty of recovering the
zeroth order term in the tree force. For a translating cluster, both
implementations of {\it Enzo} appear to lose momentum over time. 

All our tests lead to the conclusion that in order to achieve roughly
the same resolution in grid and particle codes a good rule of thumb is
to require one particle per cell in the region of interest. This is difficult to achieve for the SPH codes over shock fronts or voids and likewise harder for AMR codes to match in the centre of collapsed objects. This is reflected strongly in our results where the AMR codes largely excel at the shock tests in section~\ref{sec:shock} and the SPH codes at the gravitationally bound cluster tests in section~\ref{sec:cluster}. For a
cosmological simulation these criteria would equate to having the minimum mesh
cell size about the same size as the gravitational softening or half
the SPH search length.

As implied by the relation above, for strong shocks SPH codes require significantly more effort than AMR
codes. This is because AMR codes have the ability to add extra
resolution in regions of rapidly changing density, whereas SPH
codes can effectively only refine with the density itself, so they end up
undertaking lots of unnecessary work far from the shock where the
forces are small in regions of high but uniform density. However, for
models such as the King sphere, where AMR codes traditionally refine
using a density criterion anyway, SPH codes can achieving high resolution far more efficiently in these dense structures because
they do this very naturally by following individual particles. For AMR codes to achieve similar results, high resolution and a significant increase in computational time is required. In
these cases AMR codes would still have an advantage if it was
necessary to resolve the low density region, where the SPH codes
simply run out of particles and have to smooth over very large
volumes. 

Additional care must be taken for the parallel implementations of {\it Hydra} and {\it Enzo}. {\it Hydra}'s parallel version was not used in this paper and further testing would be needed to ensure that it performs correctly. The gravity solver in {\it Enzo} was demonstrated to produce different results in serial and parallel in the King model tests (although not in other situations) and this should be explicitly tested when running a simulation with this code. An updated gravity solver for {\it Enzo} which rectifies this problem is currently being developed. While serial runs allow much parameter space exploration, rapid code development and inter-comparison, true state-of-the-art large simulations are confined to codes that can utilise high performance computing facilities and successfully run on a large number of processors. 

To a large extent the choice of simulation code largely comes down to
the problem being attacked. For problems with large dynamic range or when
the object is rapidly translating across the volume particle based
methods are much less computationally expensive in order to achieve
the same result. Conversely, for problems where the interesting
physics is in regions of rapidly changing density rather than in high
density regions themselves, AMR codes excel thanks to their more
adaptive refinement criteria. 

The computational time required for each code to perform these tests varied greatly. Ideally, elapsed time is largely irrelevant since the required resolution and problem size should be set by the appropriate physics being tackled rather than available computer resources. However, we include Table~{\ref{table:times}} for completeness and to give the reader an indication of times involved. In Table~\ref{table:times}, the computational systems referred to are; `UF HPC' is the University of Florida's high performance computing center which uses (typically) InfiniBand-connected 2.2 GHz AMD Opteron duel cores with 6 Gb RAM. `UF Astro' is the University of Florida Astronomy's minicluster with 2.66 GHz Intel quad-cores and 4 Gb RAM. `Columbia' is the Columbia University Astronomy's cluster with gigabit-connected 2.2 GHz Opteron cores and 3 Gb RAM. `Miranda' is at the University of Durham and consists of Myrinet-connected 2.6 GHz Opteron cores with 4 Gb RAM. `Nottingham' is the Nottingham University HPC with gigabit-connected 2.2 GHz Opteron cores with 2 Gb of RAM.
 
\begin{table*}
\caption{Computational time to perform the presented tests, including the number of processors used and the type of system. Readers should note that comparing CPU times between simulations performed on disparate computers is fraught
with danger. For true runtime performance, the codes would need to be bench marked on identical systems which was beyond the scope of this paper. We therefore include this table purely for guidance and warn against over interpretation. $\dagger$ The times quoted for the static and translating cluster runs in {\it Flash} are estimated total runtimes based on the speed up attained with the latest version of the FFTW gravity solver. The original runs were performed using an early version of the hybrid FFTW gravity solver which was still under development and as such was not optimised. The original runs took place on 24 processors and runtimes for the pre-development level FFTW solver were 217 hrs 42 mins for the static cluster and 564 hrs 5mins for the translating cluster. This drastic improvements arises from a decrease in the memory requirements and improved communication patterns. A reduction in computational time in the Sedov blast test was additionally achieved by using a simplified Riemann solver that was not required to follow multiple fluids with different adiabatic indices (a feature the default configuration of {\it Flash} has implemented to model thermonuclear flashes). This reduced the memory overheads.}
\begin{tabular}{lccl}
\hline
{\bf Riemann Shock Tube: Planar} & hh:mm & no. processors & system  \\
\hline
{\it Enzo (PPM)} & 00:36 & 16 & UF HPC (Opteron) \\
{\it Enzo (Zeus)} & 04:00 & 1 & UF Astro (Intel Xeon) \\
{\it Flash} & 00:24 & 32 & Miranda (Opteron) \\
{\it Hydra} ($10^6$ particles) & 00:36  & 1 & Nottingham (Opteron)\\
{\it Gadget2} ($10^6$ particles) & 01:31 & 2 & Nottingham (Opteron) \\
\hline
{\bf Riemann Shock Tube: Oblique} & & &\\
\hline
{\it Enzo (PPM)} & 01:22 & 16 &  UF HPC  (Opteron) \\
{\it Enzo (Zeus)} & 05:06 & 8 & UF HPC (Opteron) \\
{\it Flash} & 00:26 & 45 & Miranda (Opteron)  \\
{\it Hydra} ($10^6$ particles) & 00:36 & 1 & Nottingham (Opteron) \\
{\it Gadget2} ($10^6$ particles)  & 01:30 & 2 & Nottingham (Opteron) \\
\hline
{\bf Sedov Blast Test} &  &  &  \\
\hline
{\it Enzo (PPM)} & 00:24 & 16 & UF HPC (Opteron) \\
{\it Enzo (Zeus)} & 04:41 & 6 & UF HPC (Operton) \\
{\it Flash} & 15:33 & 16 & Miranda (Opteron) \\
{\it Hydra} & 06:49 & 1 & Nottingham (Opteron) \\
{\it Gadget2} & 19:41 & 4 & Nottingham (Opteron) \\
\hline
{\bf Static Cluster Test} & &  &  \\
\hline
{\it Enzo (PPM)} & 61:50 & 1 & UF HPC (Opteron) \\
{\it Enzo (Zeus)} & 23:49 & 1 & UF HPC (Opteron) \\
{\it Flash} & 111:01 $\dagger$ & 12 & Miranda (Opteron) \\
{\it Hydra} & 37:51 & 1 & Nottingham (Opteron) \\
{\it Gadget2} & 09:48 & 4 & Nottingham (Opteron) \\
\hline
{\bf Translating Cluster Test} &  &  &  \\
\hline
{\it Enzo (PPM)}&  78:50  & 1 & Columbia (Opteron) \\
{\it Enzo (Zeus)} & 70:45 & 1 & UF Astro (Intel Xeon)\\
{\it Flash} & 374:12 $\dagger$ & 12 & Miranda (Opteron) \\
{\it Hydra} & 39:23 & 1 & Nottingham (Opteron) \\
{\it Gadget2} & 09:18 & 4 & Nottingham (Opteron) \\
\hline
\end{tabular}
\label{table:times}
\end{table*}

While this paper attempts to cover the major features of astrophysical simulations, it does comprise only four tests. Further examples would extend this paper beyond its scope (and readability), but the differences between codes cannot be fully cataloged without further testing. This paper then, is designed as a starting point for a suite of tests to be developed from which codes can be quantitatively compared for the jobs they are intended for. To assist groups wishing to run these tests on their own code and to compare new updates, we are making these results and initial conditions available on the web at {\tt http://www.astro.ufl.edu/codecomparison}.

\section*{Acknowledgements}

The authors would like to thank Alexei Kritsuk, Volker Springel \& Richard Bower for
helpful suggestions and advice and Lydia Heck for computational support. EJT acknowledges support from a
Theoretical Astrophysics Postdoctoral Fellowship from Dept. of
Astronomy/CLAS, University of Florida and both EJT and GLB acknowledge support from NSF grants
AST-05-07161, AST-05-47823, and AST-06-06959. DM acknowledges support from the EU MAGPOP Marie Curie Research and Training Network. Many simulations with
{\it Enzo} were performed at the National Center for Supercomputing
Applications and at the University of Florida High-Performance Computing
Center who also provided excellent computational support. The {\it Flash} software used in this work was in part
developed by the DOE-supported ASC / Alliance Center for Astrophysical
Thermonuclear Flashes at the University of Chicago. The {\it Hydra}
and {\it Gadget2} simulations were carried out on the Nottingham HPC
facility. The {\it Flash} simulations were carried out on the Virgo
Consortium computing facility in Durham.

\end{document}